\documentclass[12pt,preprint]{aastex}

    \newcommand{\bB}      {{\bf B}}
    \newcommand{\lamperp} {\lambda _{\perp}}

\begin{document}

\title{Coronal heating distribution due to low-frequency
wave-driven turbulence}

\author{P. Dmitruk$^1$, W. H. Matthaeus$^1$, L. J. Milano$^1$, \\
        S. Oughton$^2$, G. P. Zank$^3$ and D. J. Mullan$^1$}

\affil{
$^1$Bartol Research Institute, University of Delaware,
       Newark, DE 19716 \\
$^2$ Department of Mathematics, University College London,
       England \\
$^3$ Institute of Geophysics and Planetary Physics (IGPP),
     University of California, Riverside
         }

\email{pablo@bartol.udel.edu}

\begin{abstract}
The heating of the lower solar corona is examined using
numerical simulations and theoretical models of
magnetohydrodynamic turbulence in open magnetic
regions. A turbulent energy cascade to
small length scales perpendicular to the mean magnetic field 
can be sustained by driving with low-frequency Alfv\'en waves
reflected from mean density and magnetic field gradients. This mechanism
deposits energy efficiently in the lower corona, and we show 
that the spatial distribution of the 
heating is determined by the mean density through the
Alfv\'en speed profile. 
This provides a robust heating mechanism that 
can explain observed high coronal temperatures
and accounts for the significant heating (per unit volume) distribution
below two solar radius needed in models of the origin of the solar wind.
The obtained heating per unit mass on the other hand is much more
extended indicating that the heating on a per particle basis persists
throughout all the lower coronal region considered here.
\end{abstract}

\keywords{Sun: corona --- MHD --- turbulence}

\section{Introduction}

The mechanism by which the solar corona is heated is a fundamental
problem in astrophysics and plasma physics, and one that
remains incompletely explained.
An acceptable mechanism transports energy to the corona,
heating it to temperatures greatly in excess of chromospheric values.
This mechanism accelerates the solar wind
and establishes the boundary conditions for the entire plasma heliosphere.
There are several constraints on an acceptable coronal
heating model. Spectrometers \citep{KohlEA95,KohlEA97}
show that proton temperatures in polar coronal holes 
rise to several $10^6$\,K at
heliocentric distances of 1.5 to 2 solar radii ($R_s$)
and the bulk wind speed reaches 200\,km/s.
Remote sensing observations \citep{GrallEA96} indicate wind speeds up
to 600\,km/sec at 4\,$R_s$.
Fluid models of the solar wind
 \citep{HabbalEA95,McKenzieEA95,EvjeLeer98}
support the conclusion that wind acceleration at such low
altitudes requires heating low in the corona. 
In the fluid models an \emph{ad hoc} heating function $Q(r)$
is assumed, which represents the energy deposition per unit volume as
$Q \equiv Q_0 \exp \left[ -\frac{r - r_0}{L} \right]$,
where $r$ is heliocentric distance, $r_0=R_s\approx 7\times 10^5 $\,km,
$L$ is a prescribed dissipation lengthscale, and $Q_0$ is chosen so that the
energy flux has the required value of about
$5 \times 10^5~{\rm erg}~{\rm cm}^{-2}{\rm s}^{-1}$ \citep{WithbroeNoyes77}.
Despite the fact that there is no direct observational evidence for the 
exponential heat function, it has been  
long known to provide \citep{HolzerAxford70,KoppOrrall76,Hammer82}
rapid acceleration of the wind, and with
   $L \approx 0.2$--$ 0.5$\,$R_s$ it accounts for many observed features of
the corona \citep{McKenzieEA95,HabbalEA95,AxfordMcKenzie97}.
Although some possible sources of heating are proposed, 
in those models there is no physical mechanism for explaining this 
heating function and, being an ad hoc quantity, the exact explicit exponential
function expression is perhaps irrelevant and the main point to be remarked
is the requirement of strong heating per unit volume in the
first solar radius of the corona.
Yet other two-fluid (electrons and protons) models 
\citep{TuMarsch97,MarschTu97} have included some 
mechanisms for the heating, based on
the requirement of high frequency waves launched
from the Sun, but those high frequencies (of the order of KHz) waves
have yet remained unobserved in the coronal base where the energy is
injected. In another previous approach,
\citealp{Hollweg86,HollwegJohnson88,Isenberg90} conjectured
that waves are launched from the Sun at low frequencies but 
magnetohydrodynamic (MHD) turbulent-like cascades are assumed to 
transfer the energy to 
higher frequencies, where ion cyclotron mechanisms can be invoked to
dissipate it. This idea, recently employed by \citealp{LiEA99} and
reviewed by \citealp{Cranmer00} involve phenomenological turbulent rates 
(being called of the Kolmogorov and Kraichnan type) but two 
important problems are faced. 
First, a cascade in wave frequencies is assumed, which
implies a turbulent cascade proceeding in the parallel (to the mean
field) wavenumber direction and a linear dispersion relation to persist
between parallel wavenumber and frequency. 
MHD simulation studies show on the other hand that in 
the presence of a strong mean magnetic field, like in the corona,
the parallel wavenumber cascade is almost completely supressed 
\citep{ShebalinEA83,OughtonEA94,MattEA98} in favor of the 
perpendicular cascade. This is precisely the kind of anisotropy which is
well documented \citep{ArmstrongEA90} for the lower coronal density.

Second, the necessity of counterpropagating fluctuations to sustain
the turbulence $~$ \citep{Kraichnan65,DobroEA80} is not considered 
in the models and a WKB approach is implicitly assumed. 
In those models, the turbulent cascade was not the central
focus of study, but we believe that a useful perspective can be gained
into the heating problem by further study of this issue. 
In this present work, instead of assuming a parallel wave cascade, we describe
\emph{ab initio} the conditions under which a strong (perpendicular) 
turbulent cascade can actually be developed
and sustained by low frequency driving at the base. Reflections
from density and mean field gradients provide the necessary 
counterpropagating fluctuations to sustain the turbulence and this
effect is fully taken into account in our model. We will employ both  
direct numerical simulation and turbulent phenomenologies to obtain
the heating rate distribution. 

We show that the heating distribution profile is related to
the coronal background magnetic field and density profile, i.e.,
the heating and the density profile are not independent quantities. 
On the one hand, we find that the heating per unit volume is significant
mostly within the first solar radius, with a dissipation length scale (defined 
as the length in which the heating decays by an order of magnitude)
of a fraction of solar radius. On the other hand the heating per
unit mass varies much less dramatically and is significant through
the complete domain.
For a complete model, appropriate kinetic processes will need 
to be identified to absorb the small perpendicular scale cascaded energy.
Candidates for small scale damping are plentiful, and include oblique 
kinetic Alfv\'en waves,
collisionless reconnection and associated heating at current sheets
and nonlinear beam instabilities \citep{LeamonEA98}.
The kinetic physics of heating is beyond the scope of the present paper.
The main purpose of this work is to show that
a heating distribution emerges
naturally from a consistent MHD turbulent cascade
driven by low frequency fluctuations at the coronal base.
This result may bring us closer to understand the puzzle of
coronal heating and the origin of the solar wind.

\section{Model}

Most coronal heating models derive their energy supply from either
wave power or quasistatic field line motions,
originating in the photosphere or chromosphere.
These differ mainly in their characteristic timescales.
A general scenario for wave production in open magnetic
regions has been proposed \citep{McKenzieEA95} in which activity in
the network regions, driven by convective motions, provides a source of upward
traveling Alfv\'en waves. $~$Higher frequency waves
are easily transported upward, whereas low-frequency waves
experience greater reflection and transmit less readily
into the corona \citep{Parker65,HeinemannOlbert80,AnEA90,
ZhouMatt90,Velli93}.
However even in the limit of sharp gradients or discontinuities, there are
indications \citep{Hollweg81} that a substantial energy flux
can reach the coronal base \citep{KudohShibata99}.

Here we examine the following scenario \citep{MattEA99a}:
MHD-scale ( low-frequency ) fluctuations are excited
below the base of the corona, on transverse length scales that are
characteristic of the chromospheric network, and propagate upwards
along the mean magnetic field $\bB$.
A portion of this wave flux enters the corona.
Reflections from large-scale inhomogeneities within the lower corona
produce counter-propagating fluctuations \citep{HeinemannOlbert80,ZhouMatt90}
a situation that permits strong nonlinear MHD couplings \citep{DobroEA80}.
The couplings drive a perpendicular (to $\bB$) turbulent
cascade, producing small scale reconnection events that
couple to kinetic processes at small scales and oblique
wavevectors \citep{LeamonEA98}.

In our model, we simulate the propagation of low-frequency MHD
fluctuations in a medium whose properties are chosen to represent
an open magnetic field region (like a polar coronal hole) in the lower corona.
We consider a volume section which extends in the radial
direction a distance $L_s$ of order $2 R_s$ above the surface.
A sketch of such a region is shown in Fig. 1.

Our goal is to compute the dynamics of transverse fluctuations of
magnetic field $\bf b$ and velocity $\bf v$
as influenced by specified large-scale inhomogeneities (mean magnetic
field and density gradients) as well as by local nonlinear couplings.
To do this, in a medium where there is a strong mean magnetic field,
we use the reduced MHD (RMHD) approximation
\citep{Strauss76,Mont82,ZankMatt92}.
This approach is appropriate for fluctuations in a nearly incompressible 
plasma, where the plasma beta (ratio of thermal to magnetic pressure) 
is low \citep{ZankMatt92}. 
We employ the Elsasser variables $\bf z_{\pm}$ = $\bf v$ $\pm$ $\bf b$
(where the magnetic field is written 
in units of ${\bf b}/\sqrt{4\pi\rho}$, with $\rho$=density)
to represent downward (+) and upward (-) propagating fluctuations.
Fluctuations depend on the radial coordinate $r$, the transverse spatial
coordinates (including an areal expansion factor $A(r)$) 
and on time $t$. Mean quantities like density $\rho$ and 
magnetic field ${\bf B}$ are assumed to depend only on $r$.
Consistency with the RMHD assumptions \citep{ZankMatt92} requires
$\partial_r \ll \nabla _{\perp}$, $V_A \gg z_{\pm}$,
and $\nabla _{\perp} \cdot {\bf z}_{\pm}=0 $, 
where ${\bf V_A} = {\bf B} / \sqrt{4\pi\rho}$ is the Alfven speed
which gives the propagation velocity of fluctuations in the linear limit.

In the presence of a non-uniform $V_A$, the nonlinear RMHD equations can 
be written \citep{ZhouMatt90} as
\begin{eqnarray}
  \frac{\partial{\bf z}_-}{\partial t}
    + V_A \frac{\partial {\bf z}_-}{\partial r} & = &
     -R_1 {\bf z}_+ + R_2 {\bf z}_-
   - {\bf z}_+ \cdot \nabla_\perp {\bf z}_-
   + \eta \nabla_\perp^2 {\bf z}_-
\nonumber \\
  \frac{\partial{\bf z}_+}{\partial t}
    - V_A \frac{\partial {\bf z}_+}{\partial r} & = &
     R_1 {\bf z}_- - R_2 {\bf z}_+
   - {\bf z}_- \cdot \nabla_\perp {\bf z}_+
   + \eta \nabla_\perp^2 {\bf z}_+
 \label{eq:dzdt}
\end{eqnarray}
where $\eta$ is the resistivity (assumed equal to the viscosity),
and $R_1 (r), R_2(r)$ are specified reflection rates due to the
inhomogeneous $V_A (r)$.
For a symmetric radial geometry ($A(r)=r^2$), 
$B=B_0 (r_0/r)^2$ (with $B_0$ the magnetic
field at $r_0=R_s$), the reflection rates are
$ R_1 (r) = \frac{1}{2} \frac{d V_A}{dr} $
and $ R_2 (r) = \frac{1}{2} \frac{d V_A}{dr} + \frac{V_A}{r} $.
In the case of a general geometry with cross section area $A(r)$,
where $A(r)=A_0~B_0/B(r)$, the reflection rates are
$ R_1(r) = \frac{1}{2} \frac{d V_A}{dr}$ and
$ R_2(r) = \frac{1}{2} \frac{d V_A}{dr} + \frac{1}{2} \frac{V_A}{A}
           \frac{dA}{dr} $. The previous expressions are valid for
a section with transverse dimensions small compared to the longitudinal size
and close to the polar region in the Sun (dependence of the mean field
$B$ with the latitudinal angle should be assumed to maintain the 
free divergence condition, $\nabla \cdot {\bf B} = 0$).

A Chebyshev-Fourier representation is chosen to solve the equations
(\ref{eq:dzdt}) using a direct numerical simulation of the pseudospectral
type. Transverse coordinates are periodic, while non periodic
boundary conditions are imposed on the bottom and top sections of the
volume.
In order to model the input of
velocity fluctuations at the base of the coronal
domain, we refer to observations of non-thermal velocity amplitudes in the
upper transition region of the Sun. These amplitudes are reported to
range from 20 to 55 km/sec \citep{ChaeEA98,DoyleEA98,HasslerEA90}.
The system of nonlinear equations (\ref{eq:dzdt}) is driven
by imposing a time dependent oscillatory
pattern of velocity and magnetic field at the bottom surface,
with a transverse lengthscale corresponding to
supergranule diameters of 13-32 Mm at the coronal base \citep{HagenaarEA97}
and the typical inter-network spacing
of 30 Mm \citep{AxfordMcKenzie97}. 
Transverse dimensions of the simulation box are chosen to include about
4 supergranules. In the bottom panel of Fig. 2 the boundary conditions imposed
on the form of waves at the bottom of the domain are shown. 
The top panel of Fig. 2 
shows a cross section of the volume above in the coronal simulation 
(corresponding to one of the simulations to be described below). 
Numerical resolution of $128 \times 128 \times 33$ Fourier+Chebyshev 
grid points are employed in this simulation, for a macroscopic Reynolds 
number of order $600$. 
The fluctuating fields are highly structured
on the transverse directions as shown by concentrations of current density
in the form of sheet-like small scale structures.

Driving is at
a low frequency of $ f = 0.6 / t_A $,
where   $ t_A =  R_s/ V_{A_0}$ is a vertical Alfv\'en crossing time based
on the Alfv\'en velocity at the base and a distance of a solar radius.
This correspond to driving wave periods of the order of 1000 sec. 

Open boundary conditions are assumed at the top surface.
If reflections were absent, this would
lead to a wavetrain of upward propagating fluctuations that would
escape continuously from the upper boundary without depositing any
energy in the system. The system would dynamically relax toward a 
maximal cross helicity state \citep{DobroEA80,GrappinEA83,TingEA86}
in which downward propagating fluctuations vanish and there would be
no incompressible MHD nonlinearities and hence no cascade.
However, the first two terms on the right hand sides of
Eqs.~(\ref{eq:dzdt}) are associated with gradients of the wave speed, and these
lead to partial reflection of the excited waves.
This results in a state with both upward and downward type fluctuations
and thus, sustained turbulence. 
We have presented in a previous numerical study \citep{DmitrukEA01a} 
the conditions under which an efficient MHD turbulent perpendicular 
cascade can be sustained; what is required is occurrence of reflections and 
non-propagating structures (controlled by the type of boundary conditions 
imposed). 
The efficiency of this mechanism on dissipating the injected energy
has been demonstrated by both phenomenologies 
\citep{MattEA99a,DmitrukEA01b} as well as direct numerical simulations 
\citep{DmitrukEA01a,OughtonEA01}.
Efficiencies (ratio of turbulent dissipation rate by
the energy injection rate) averaging between 10 and 40 \% can be obtained. 
Even very weak reflections are capable
of sustaining the turbulence and provide effective dissipation. 

In the next section we obtain the heating
distribution profiles resulting from this turbulent state. 

\section{Heating profiles}

We performed numerical simulations employing several different mean magnetic
field and density radial profiles. The values at the bottom boundary
\citep{AxfordMcKenzie97,FeldmanEA97}
have been fixed to $B_0 = 9$ Gauss, and a number density of
$n_0 = 3.2 \times 10^8~{\rm cm}^{-3}$, which gives an Alfven speed at
the base of $1100$ km/s. With the assumed amplitude of 
fluctuations of the order of $20-55$ km/s, this gives an average energy
input flux per unit area, $F_{A_0} \sim \rho _0 V_{A_0} <z_-^2 - z_+^2>$
of order $5 \times 10^5 {\rm erg}~{\rm cm}^{-2}~{\rm s}^{-1}$ corresponding
to the \citealp{WithbroeNoyes77} value. These values have to be taken as 
order of magnitude quantities but a different choice would
not invalidate the general argument we want to present here. 
In our first simulation we consider a symmetric radial mean magnetic
field profile $B(r)=B_0 (R_s/r)^2$.
The mean density follows from the hydrostatic 
equilibrium approximation
\citep{AnEA90,Velli93,NakariakovEA00}:
$dp/dr$ = -$m_i n g$, with $p$ = thermal pressure, $n$ = number density,
$m_i$ the proton mass and $g$ = gravity. 
In an isothermal corona, this equation
takes the form $dn/dr = -\alpha_T~n R_s/r^2$ where 
$\alpha _T= \mu GM/R_g T R_s$ ($\mu$ mean molecular weight, $G$ gravitational
constant, $M$ solar mass, $R_g$ gas constant and $T$ temperature).
The quantity $\alpha _T \approx 12$ for the Sun and a fully ionized hydrogen 
coronal plasma at $T=10^6$ K. 
The number density
distribution is then $n(r)$ = $n_0$ exp[$\alpha _T(R_s/r ~-~1)$]
and the density is $\rho (r) = m_i n(r)$.
We consider two such density profiles, shown in the top left panel of Fig. 3
with continuous and dotted lines, corresponding to values of
$\alpha _T = 12$ and $\alpha _T = 6$ (representing a hotter corona).
This gives two different density profiles, with the hotter corona 
having the shallower profile. The corresponding
Alfven speed profiles $V_A(r)=B(r)/\sqrt{4\pi\rho(r)}$
are shown on the top right panel of Fig. 3. 
When the system reaches a statistically steady state the turbulent 
dissipation rate $\epsilon$ (equal to the cascade rate)
is computed from the fluctuating fields 
${\bf z_-}$, ${\bf z_+}$
from which the heating per unit volume is obtained as 
$Q = \rho \epsilon$. 

This quantity is plotted as a function of $r$ in the bottom left 
panel of Fig. 3. It is apparent that the dissipation
per unit volume is peaked near the model coronal base. It is important to
note that $Q(r)$ decreases faster for the model with the steeper
density profile. 
The radial distance $\Delta r$ at which $Q$ decreases by an order
of magnitude with respect to the coronal base value $Q_0$
is $0.36 R_s$ and $1.1 R_s$ for the two profiles in Fig. 3. This would be
the analog of the dissipation length in the ad-hoc exponential heating 
distribution profiles. The bottom right panel on Fig. 3
shows the heating per unit mass, a useful quantity to consider
the heating in a per particle basis. 
Over our entire domain this quantity maintains a relatively constant value.
Extended heating per unit mass fits the apparent requirement of a hot
(non adiabatic) solar wind and has also emerged in empirical
models of radiation loss and other heating models 
\citep{MullanCheng94,Gibson73}.

A third simulation has been performed employing a super
radial geometry, where the mean magnetic field profile has
been taken as 
\begin{equation}
B = 1.5 \left[ (f_{max}-1) \left(\frac{R_s}{r}\right)^{3.5} +
                 \left(\frac{R_s}{r}\right)^2 \right] ~{\rm Gauss}
\end{equation}
Profiles of this type have been considered in \citealp{Hollweg00}
and similar super expansion profiles have been used by 
\citealp{McKenzieEA95,EvjeLeer98,LiEA99} in their
solar wind models.
We have taken $f_{max}=6$ which gives a boundary value of $B_0=9$ Gauss. 
An observationally based number density is assumed in this case
\citep{FeldmanEA97},
\begin{equation}
n(r)= 3.2 \times 10^8 \left(\frac{R_s}{r}\right)^{15.6} +
      2.5 \times 10^6 \left(\frac{R_s}{r}\right)^{3.76} ~{\rm cm}^{-3}
\end{equation}
The density profile and resulting Alfven speed profile are shown in
the top panels of Fig. 3 (dashed line). 
The obtained heating per unit volume and heating per unit mass 
are shown in the bottom panels of Fig. 3 (dashed line).
The result confirms the tendency shown for the earlier cases:
this profile, with the fastest decreasing density, also has the
fastest decreasing heating per unit volume. 
The decay length for this case is $\Delta r = 0.2 R_s$. 
The heating per unit mass remains at a relatively constant value.

The relevant results to be extracted from these simulations for different
Alfven speed profiles is that the heating per unit volume has
significant values only within 
the first solar radius of the lower corona and that
the radial profile of the heating is controlled by
the radial profile of the density. As it will be more clear from 
the phenomenological analysis that follows in the next section, it is 
actually through the Alfven speed profile and its gradient 
that this behavior is manifested. 

Whether the radial profile of the heating per unit volume is 
exactly exponential or not is, 
we believe, irrelevant and we conjecture that a steep
heating profile, like the ones we obtained here, would work
as well to explain the fast acceleration of the solar wind. 

Fig. 3 contains the essence of the present results,
that a reasonable and efficient heating can be obtained from
the MHD perpendicular turbulent cascade driven by low frequency
waves injected from the bottom. The properties of this heating are controlled 
by both the amount of energy injected into the system and the
background profiles of magnetic field and mean density. To further
illustrate this idea and, in some limiting case, to obtain an
analytical expression for the heating profile in the low region
of the volume considered here, we present in the following section a
phenomenological approach to the previous model Eqs (\ref{eq:dzdt}).

\section{Heating profile from a phenomenological model}

A nonlinear phenomenology \citep{DmitrukEA01b} simplifies the study of
the wave-driven RMHD heating model.
We modify Eqs.~(\ref{eq:dzdt}) as follows:
the independent variables, now designated as $Z_\pm(r)$, are treated as
one dimensional wave amplitudes with the same linear transport terms
(LHS and reflection terms) as in Eqs.~(\ref{eq:dzdt}),
but with the nonlinear and dissipative terms on the RHS
replaced by the nonlinear models $Z_\mp|Z_\pm| / 2 \lamperp(r) $.
This model entirely suppresses the transverse structure,
with the strength of the nonlinear cascade effects represented through
a single correlation lengthscale $\lamperp(r)$.
Previous investigations \citep{DmitrukEA01b}
have shown that this phenomenology portrays
many of the same physical effects as wave-driven models
based upon the full RMHD equations \citep{DmitrukEA01a,OughtonEA01}.
We expect this model to correspond to high Reynolds number fully
developed RMHD turbulence since the amplitudes $Z_-$ and $Z_+$
are strongly coupled in the adopted forms for the nonlinear terms.
We extracted a heating function $Q(r)$ from
a numerical solution to this phenomenological model,
using the same parameters as those employed in
the full nonlinear RMHD simulations and
assuming a correlation lengthscale $\lamperp(r)$
linearly increasing with radius,
with a value at the coronal base set to correspond to
the inter-network spacing of 30 Mm.
The derived $Q(r)$ is illustrated in Fig. 4 as the thin line.
The direct numerical solution of the previous section
is shown as the thick line. 
This corresponds to the simpler hydrostatic density and $1/r^2$
mean magnetic field profile (similar results are obtained for
the super expansion geometry). Despite some differences on the far
region, where $dV_A/dr$ approaches $0$ and the heating is very
low, the agreement is pretty good, considering the extreme simplification
involved in the phenomenological treatment. Of course, the phenomenology
would not be useful for studies of the turbulent cascades and spectra,
but it gives the right order of magnitude for the averaged quantities
such as the dissipation (heating), when this quantity is not extremely low. 
The interesting issue of how the 
correlation length and the strength of the cascade
depend on the external parameters warrants further study comparing
the phenomenology with the full non linear direct numerical simulation.

In Fig. 5 we explore the phenomenological heat function $Q$ obtained
by varying the correlation lengthscale
at the coronal base, with $\lamperp=1$ corresponding to 30 Mm.
The limit $\lamperp \to 0$ corresponds to
stronger local turbulence since the
implied eddy turnover time $\lamperp / z_-$ becomes much
smaller than the wave transit time $R_s/V_A$.
Fig. 5 shows that $Q$ increases
as $\lamperp \to 0$, while maintaining the fast radially decreasing behavior.

In this limit, an analytical asymptotic expression can be obtained.
Although limited in its validity, this analysis serves to illustrate
better the properties of the heating in this model.
We begin by
formulating a flux balance equation, based upon Eq.(\ref{eq:dzdt}).
This takes the form $ dF/dr = -A(r) Q$, where 
the net upward energy flux $ F = \rho A~ V_A (Z_-^2-Z_+^2)$ 
and $A$ is the cross section area.
Note that in geometrical optics (or ``WKB'' theory for noninteracting waves)
\citep{Jacques77} the flux balance formally requires that $Q\equiv 0$. 
When fluctuations are damped, $Q \ne 0$ \citep{Hollweg86}.
Here the crucial relation that connects the heating to the 
Alfv\'en speed profile emerges by identifying $Q$ with a physically correct
turbulent dissipation rate per unit volume at distance $r$, 
namely $ Q = \rho ~\epsilon$,
where the dissipation rate per unit mass $\epsilon$ is 
expressed in the phenomenological model as
$ \epsilon \approx (Z^2_+|Z_-|+Z^2_-|Z_+|)/\lamperp$.
We now proceed to form an asymptotic strong turbulence limit as follows. 
Letting $\lamperp  \to 0$,
one can conclude \citep{DmitrukEA01b} that the downward flux is small,
  $ |Z_+| \ll |Z_-| $,
and that in a steady state this limit is characterized by
  $ |Z_+| / \lamperp = |dV_A/dr|$.  Thus
  $ \epsilon \approx Z_-^2|Z_+| / \lamperp $
and we obtain the asymptotic result that the dissipation per unit length 
can be related to the Alfv\'en speed profile and the energy flux:
  $ A~Q \approx \rho A~ Z_-^2|dV_A/dr| \approx V_A^{-1}|dV_A/dr| F$.
Thus, the asymptotic flux balance equation can now be written as
  $ dF/dr \approx  - V_A^{-1} |dV_A/dr| F$,
whose solution is
\begin{equation}
F \approx \cases{
    F_0 ~ V_{A_0}/V_A  &if $r < r_m$ \cr
    F_0 ~ (V_{A_0}/V_{A_m})^2~ 
     V_A/V_{A_0} &if $r > r_m$ \cr}
\end{equation}
where $F_0$ is the net flux at the base and $r_m$ is the radial distance
at which $V_A$ reaches its maximum value $V_{A_m}$. 
 
This allows us finally to
obtain an explicit expression for the heating deposition
\begin{equation}
Q(r) \approx \cases{
    F_{A_0}~(A_0/A)~|dV_A/dr|~
    (V_{A_0}/V_A^2)  &if $r < r_m$ \cr
    F_{A_0}~(A_0/A)~|dV_A/dr|~
    (V_{A_0}/V_{A_m}^2) &if $r > r_m$ \cr}
 \label{eq:Qresult}
\end{equation}
where $F_{A_0}=F_0/A_0$ is the energy flux per unit area at the base
(i.e, the \citealp{WithbroeNoyes77} value).

For the particular case of the $A=r^2$ geometry and the 
isothermal atmosphere used before, with $r_m \approx 3 R_s$,
this relation predicts the top dot dashed curve of Fig. 5. 
As seen in this figure, the phenomenological solutions for different $\lamperp$
values approach this asymptotic case as $\lamperp$ tends to 0. Already
for $\lamperp = 1/10$ the asymptotic solution would be a good approximation.
This corresponds to correlation length scales of order 3000 km. 
Although much less than the typical inter network distances,
observations do not rule out correlation scales of this magnitude.
Eq. (\ref{eq:Qresult}) also shows that in this limiting case 
the value of the heating goes to $0$ at points where 
$dV_A/dr = 0$, a limitation which
is only apparent at corresponding very low values of $Q$ in the
numerical solution (a similar result applies for the super expansion
profile presented before, where the point at which $dV_A/dr=0$
is $r_m \approx 1.6 R_s$).

For this asymptotic limit, eq. (\ref{eq:Qresult}) 
explicitly shows that the spatial distribution of turbulent heating 
is directly connected to the radial profile of the Alfv\'en speed
and its absolute value depends on the energy input flux. 
This illustrates the connection already pointed out through the
direct numerical simulations results.

\section{Conclusions}

We have presented a model for MHD turbulence driven by low-frequency waves
imposed at the lower boundary. 
We applied this model to an open region in the
lower corona, adopting reasonable profiles
for magnetic field and density. Both direct numerical simulations
and phenomenological studies have been performed and compared.
Results show that turbulence can be sustained, due to reflection of
upward traveling waves by Alfv\'en speed gradients,
producing a perpendicular turbulent cascade and efficient heating.
The energy dissipation per unit volume
is significant mostly in the lower part of the model
corona with a dissipation length scale of a fraction of solar radius.
The obtained heating profile $Q$ is
similar to profiles commonly used in solar wind
acceleration models.
However the more general result emerges, 
for strong turbulence, that the heat function 
$Q(r)$
should be related to the Alfv\'en speed profile
$V_A(r)$. This shows that 
the heating distribution is determined
by the large-scale magnetic field and density profiles.
In this way one sees that the 
scale-height of the heating per unit volume
is directly related to the density scale-height,
thus explaining the spatial confinement of $Q$ as a consequence of the
rapid density decrease in the lower corona.
However, the deposition of
heat per unit mass $Q/\rho$ is much
more extended, perhaps well beyond the lower coronal region considered
in the present model. These conclusions do not
rule out other models, such as direct cyclotron absorption of high-frequency
waves \citep{AxfordMcKenzie97,TuMarsch97} but they imply that it is
possible to provide the proper kind of heating per unit
volume using a low-frequency wave-driven MHD turbulence
mechanism.

Research supported by NASA (NAG5-7164, NAG5-8134 SEC Theory Program),
NSF (ATM-9713595, ATM-0296113), and UK PPARC (PPA/G/S/1999/00059).

\newpage

\noindent
Fig. 1 --
Sketch of the coronal region considered in this model

\noindent
Fig. 2 --
In the bottom panel the wave forcing applied at the bottom
boundary is illustrated with the imposed transverse fluctuating
field ${\bf z_-}$ (arrows) overlaying the current density (grayscale). 
In the top panel a cross section at a fraction $0.2 R_s$ above the base is
shown with transverse fluctuating field (arrows) over the current 
density (grayscale) obtained in the numerical simulation. Note
the small scale structure formation.

\noindent
Fig. 3 --
The panel at the top left shows three different number density
profiles assumed for the simulations. The continuous and dotted
line correspond to the hydrostatic model for different values
of the density scale height. The dashed line correspond to an
observationally based density (see text). The top right panel
is the corresponding Alfven speed profile, with $1/r^2$ geometry
for the hydrostatic density and super expansion geometry for
the observational density. The bottom left panel shows the result
of the heating per unit volume from the simulations. The bottom right
panel is the heating per unit mass.

\noindent
Fig. 4 --
Comparison between the heating per unit volume obtained with the
direct numerical simulation of Eqs.(1) (thick line) and the 
numerical solution of the phenomenological model
(thin line), for the case of an hydrostatic density profile
and $1/r^2$ geometry. 

\noindent
Fig. 5 --
Different solutions for the heating per unit volume 
of the phenomenological model (thin lines) for values of the 
correlation length $\lamperp=1, 0.5, 0.1, 0.02$ 
(in units of the inter network length) which approach
the asymptotic solution when $\lamperp \rightarrow 0$ 
(Eq.(5) in the text) shown as the dot dashed line.

\begin{figure}
\plotone{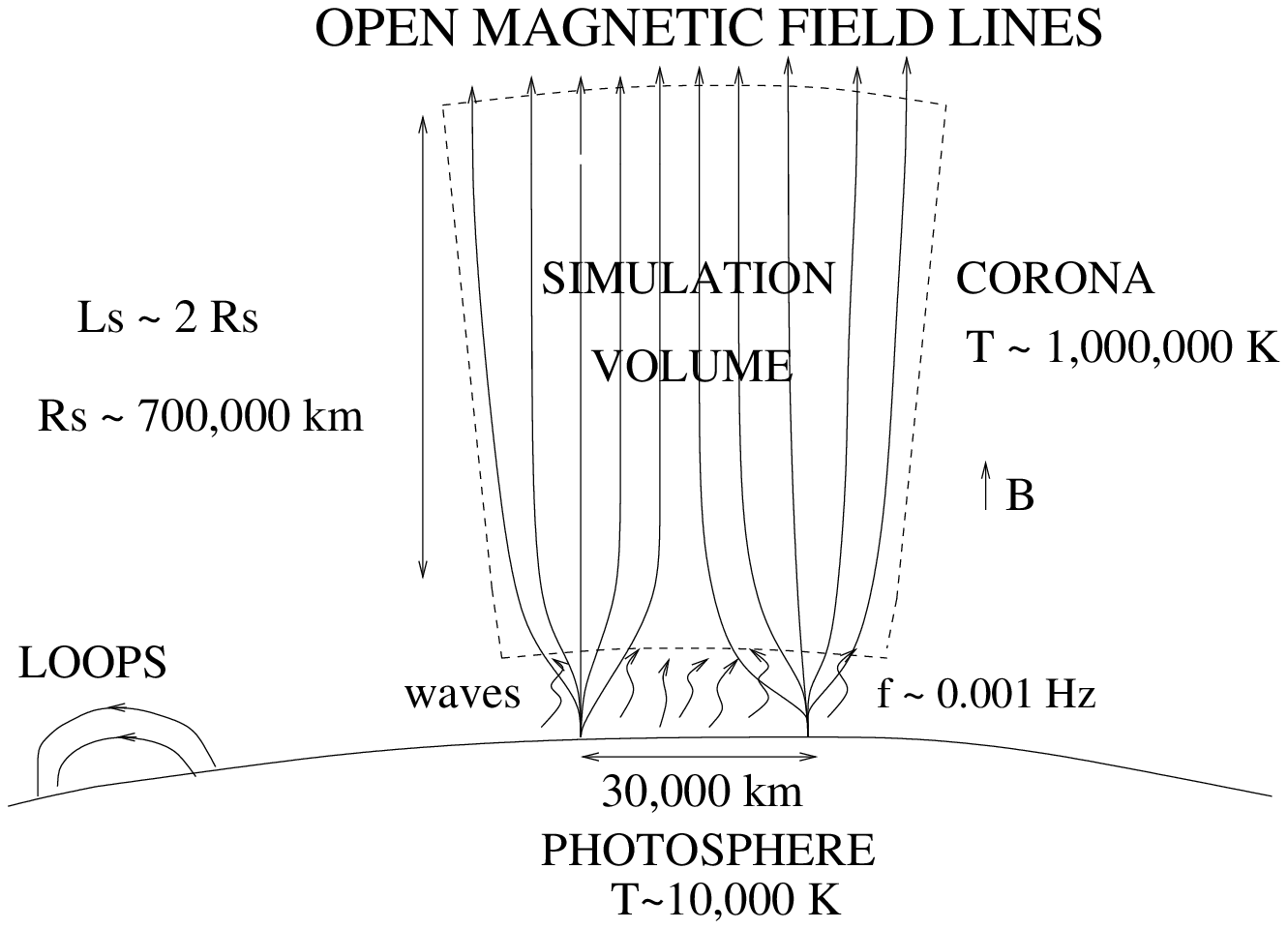}
\caption{}
\end{figure} 
\begin{figure}
\plotone{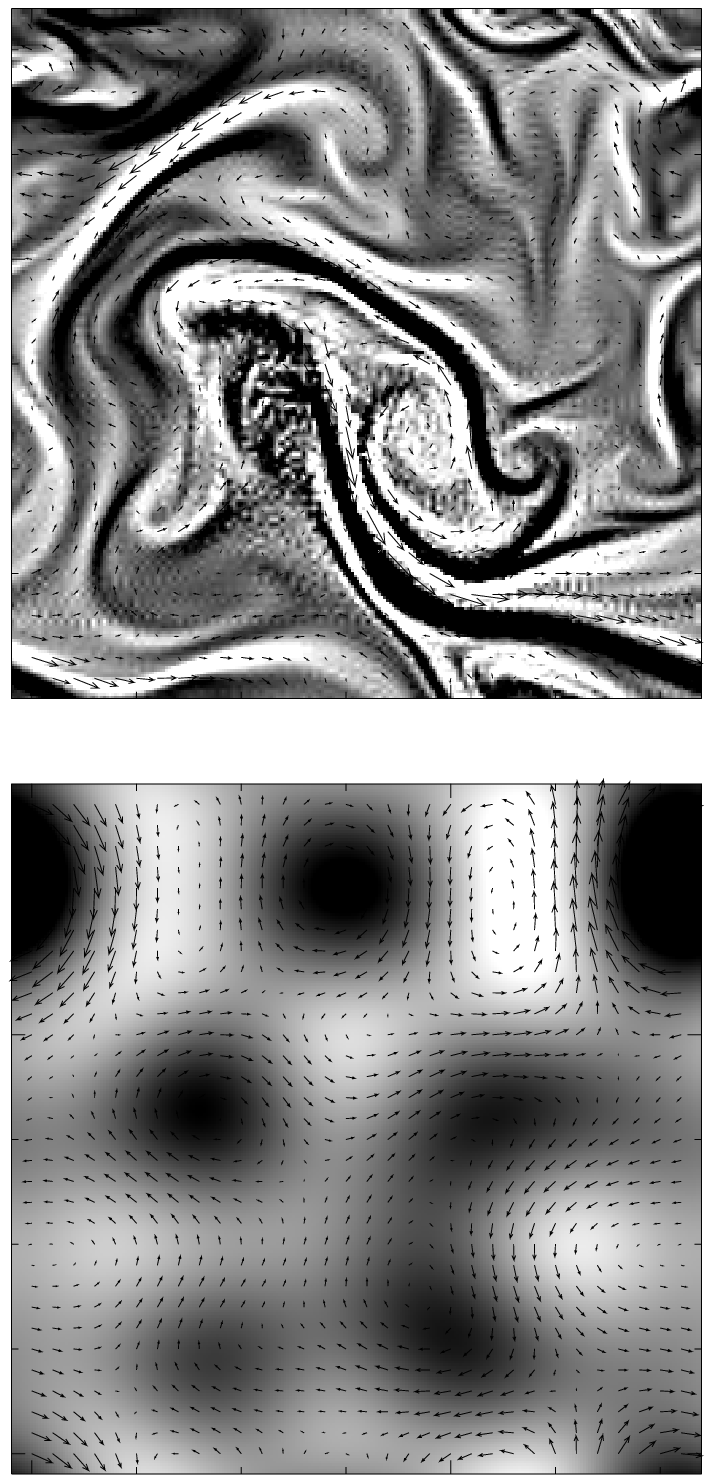}
\caption{} 
\end{figure} 
\begin{figure}
\plotone{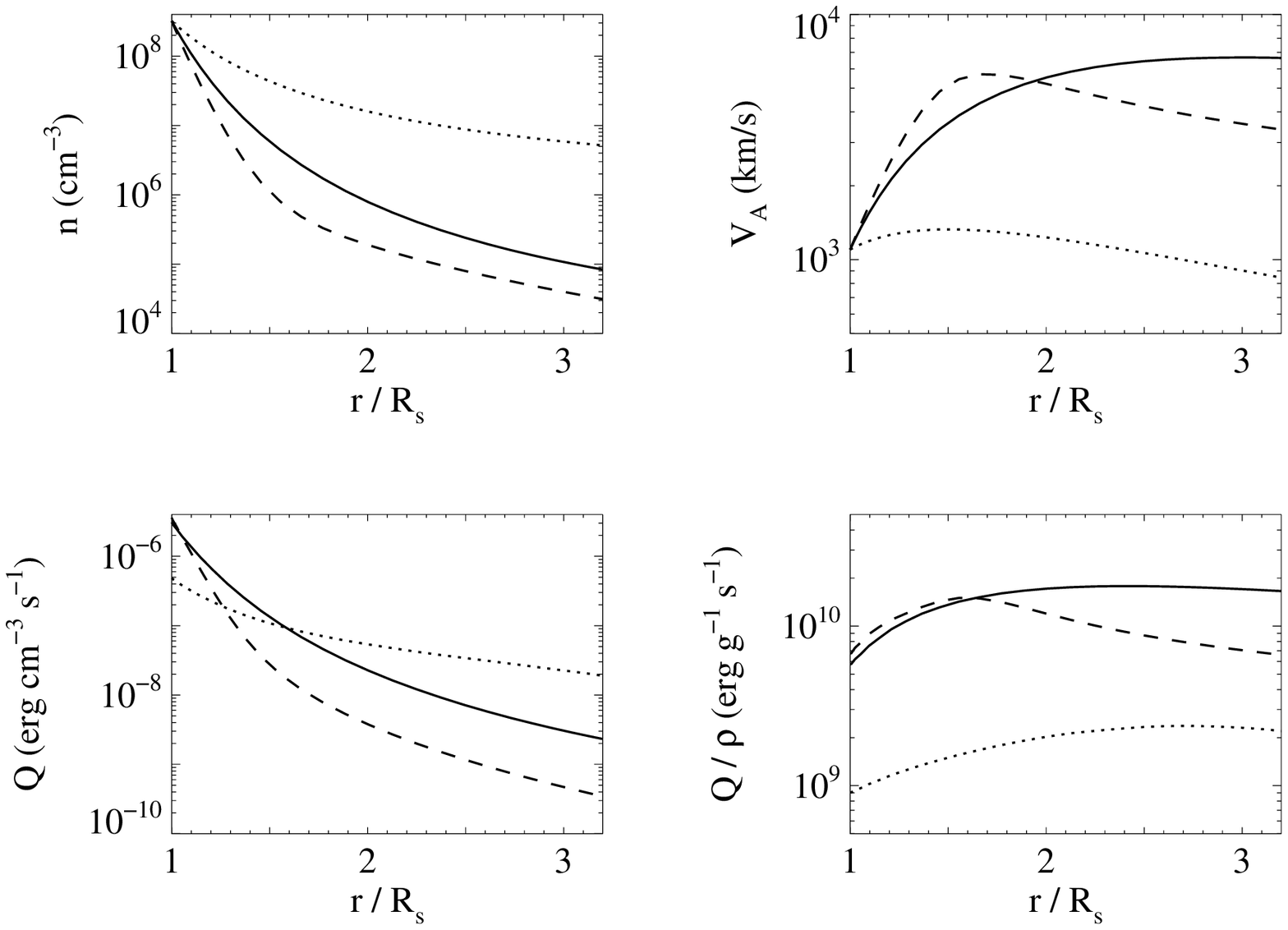}
\caption{} 
\end{figure} 
\begin{figure}
\plotone{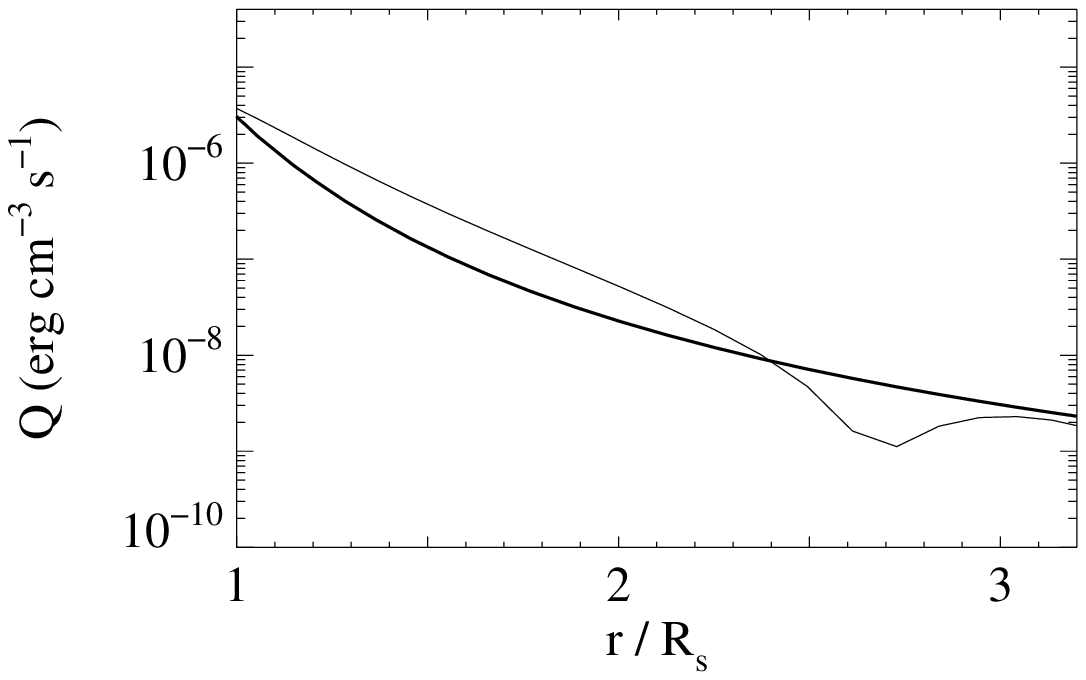}
\caption{} 
\end{figure} 
\begin{figure}
\plotone{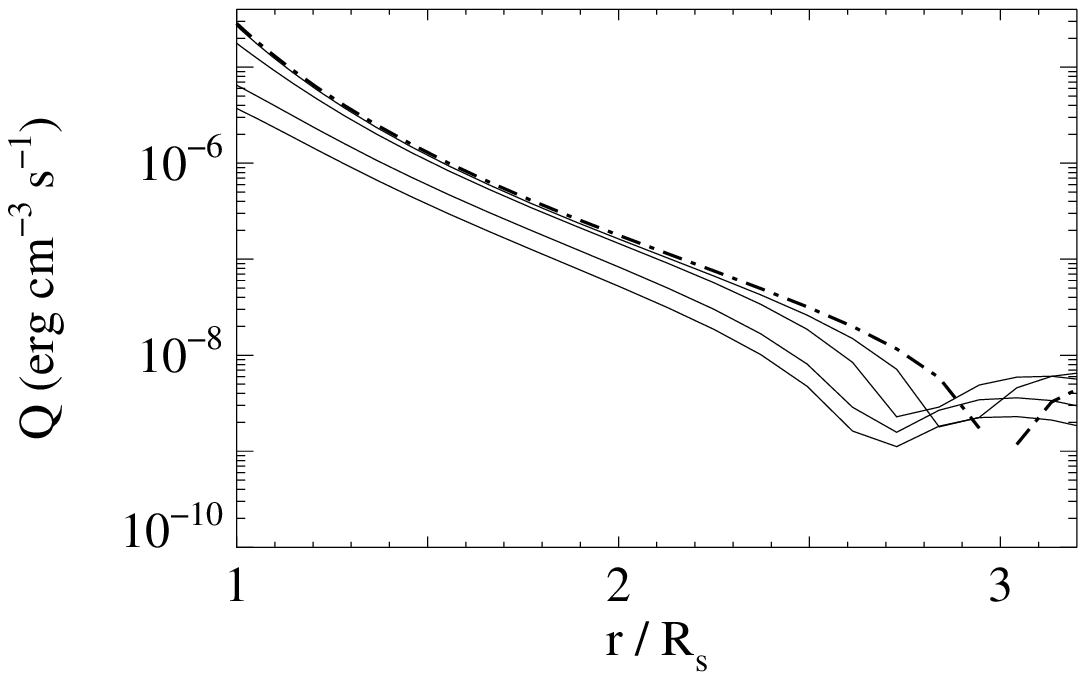}
\caption{} 
\end{figure} 

\end{document}